\documentclass{jac}
\usepackage[utf8]{inputenc}
\usepackage{cyrillic}
\usepackage[T2A,T1]{fontenc}
\newcommand\saintcyr[1]{{\fontencoding{T2A}\selectfont #1}}
\usepackage{amsmath,amssymb,graphicx}
\DeclareMathOperator{\KS}{\mathrm{K}}
\let\ge=\geqslant
\let\le=\leqslant
\begin{document}
\title{Decomposition Complexity}
\author
{A.~Shen}{Alexander Shen}
\address
{LIF Marseille, CNRS \& University Aix--Marseille;
on leave from IITP RAS, Moscow}
\email{Alexander.Shen@lif.univ-mrs.fr}
\urladdr{http://www.lif.univ-mrs.fr/\~\relax ashen}
\thanks{Author is grateful to V.V.~Podolskii, A.E.~Romashchenko,
and ESCAPE team in general for useful
discussions, and to the reviewers for very helpful
comments. The paper was supported in part by NAFIT ANR
-08-EMER-008-01 and RFBR 09-01-00709-a grants.}

\begin{abstract}\noindent
We consider a problem of decomposition of a ternary function
into a composition of binary ones from the viewpoint of
communication complexity and algorithmic information theory as
well as some applications to cellular automata.
\end{abstract}

\maketitle

\section{Introduction}

The 13th Hilbert problem asks whether all functions can be
represented as compositions of binary functions.  This question
can be understood in different ways. Initially Hilbert was interested in
a specific function (roots of a polynomial as function of its coefficients).
Kolmogorov and Arnold (see~\cite{hilbert-positive}) gave kind 
of a positive answer for
continuous functions proving that any continuous function of several
real arguments can be represented as a composition of continuous
unary functions and addition (a binary function). On the other
hand, for differentiable functions negative answer was obtained
by Vituschkin. Later Kolmogorov interpreted this result in terms
of information theory (see~\cite{hilbert-negative}): the
decomposition is impossible since we have ``much more'' ternary
functions than compositions of binary ones.
In a discrete setting this information-theoretic argument
was used by Hansen, Lachish and
Miltersen (\cite{hansen-lachish-miltersen}. We consider
similar questions in a (slightly) different
setting.

Let us start with a simple decomposition problem. An input (say,
a binary string) is divided into three parts $x$, $y$ and $z$.
We want to represent $T(x,y,z)$ (for some function $T$) 
as a composition of three binary
functions:
        $$
T(x,y,z)=t(a(x,y),b(y,z)).
        $$
In other words, we want to compute $T(x,y,z)$ under the
following restrictions:

\begin{figure}
\begin{center}
       \includegraphics[scale=0.85]{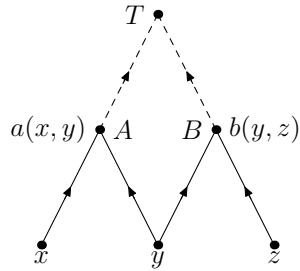}
\end{center}
\caption{Information transmission for the decomposition.}
\label{abc}
\end{figure}

\noindent
node $A$ gets $x$ and $y$ and computes some function $a(x,y)$;
node $B$ gets $y$ and $z$ and computes some function $b(y,z)$;
finally, the output node $T$ gets $a(x,y)$ and $b(y,z)$ and
should compute $T(x,y,z)$.

The two upper channels have limited capacity; the question is
how much capacity is needed to make such a decomposition
possible. If $a$- and $b$-channels are wide enough, we may
transmit all the available information, i.e., let
$a(x,y)=\langle x,y\rangle$ and $b(y,z)=\langle y,z\rangle$.
Even better, we can split $y$ in an arbitrary proportion and
send one part with $x$ and the other one with $z$.

Is it possible to use less capacity? The answer evidently
depends on the function $T$. If, say, $T(x,y,z)$ is \texttt{xor}
of all bits in $x$, $y$ and $z$, one bit for $a$- and $b$-values
is enough. However, for other functions $T$ it is not the case,
as we see below.

In the sequel we prove different lower bounds for the necessary
capacity of two upper channels in different settings; then we
consider related questions in the framework of multi-source
algorithmic information theory~\cite{multisource}).

Before going into details, let us note that the definition of
communication complexity can be reformulated in similar terms:
one-round communication
complexity corresponds to the network
       
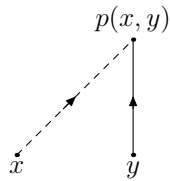
\begin{center}
\includegraphics[scale=0.85]{decomp-2.mps}
       \end{center}
\noindent
(dotted line indicates channel of limited capacity) while
two-rounds communication complexity corresponds to the network
       
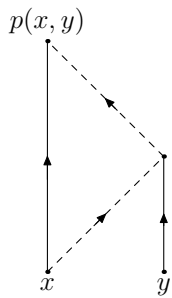
\begin{center}
\includegraphics[scale=0.85]{decomp-3.mps}
       \end{center}
etc. Another related setting that appears in communication
complexity theory: three inputs $x,y,z$ are distributed between
three participants; one knows $x$ and $y$, the other knows $y$
and $z$, the third one knows $x$ and $z$; all three participants
send their messages to the fourth one who should compute
$T(x,y,z)$ based on their messages (see~\cite{nisan}).

One can naturally define communication complexity for other
networks (we select some channels and count the bits that go
through these channels).

\section{Decomposition complexity}\label{sec:communication}

Now let us give formal definitions.
Let $T=T(x,y,z)$ be a function defined on $\mathbb{B}^p\times
\mathbb{B}^q\times \mathbb{B}^r$ (here $\mathbb{B}^k$ is the set
of $k$-bit binary strings) whose values belong to some set $M$. We say
that \emph{decomposition complexity} of $T$ does not exceed $n$
if there exist $u+v\le n$ and functions $a\colon
\mathbb{B}^p\times\mathbb{B}^q\to\mathbb{B}^u$, $b\colon
\mathbb{B}^q\times\mathbb{B}^r\to\mathbb{B}^v$ and $t\colon
\mathbb{B}^u\times\mathbb{B}^v\to M$ such that
       $$
T(x,y,z)=t(a(x,y),b(y,z))
       $$
for all $x\in \mathbb{B}^p$, $y\in \mathbb{B}^q$, $z\in
\mathbb{B}^r$. (As in communication complexity, we take into
account the total number of bits transmitted via both restricted
links. More detailed analysis could consider $u$ and $v$
separately.)

\subsection{General upper and lower bounds}

Since the logarithm of the image cardinality is an evident lower
bound for decomposition complexity, it is natural to consider
\emph{predicates} $T$ (so this lower bound is trivial).  This makes
our setting different from \cite{hansen-lachish-miltersen} where 
all the arguments and values have the same size. However,
the same simple counting argument can be used to provide
worst-case lower bounds for arbitrary functions.

\begin{theorem}
\label{th:general}
\textup{\textbf{(Upper bounds)}}~Complexity of any function does not exceed
$n=p+q+r$; complexity of any predicate does not exceed $2^r+r$
as well as $2^p+p$.

\textup{\textbf{(Lower bound)}}~If $p$ and $r$ are not too small
\textup(at least $\log n+O(1)$\textup), then there exists a
predicate with decomposition complexity $n-O(1)$.

\end{theorem}

The second statement shows that the upper bounds provided by the
first one are rather tight.

\proof (Upper bounds)~For the first bound one can let, say,
$a(x,y)=\langle x,y\rangle$ and $b(y,z)=z$. (One can also
split $y$ between $a$ and $b$ in an arbitrary proportion.)

For the second bound: for each $x,y$ the predicate $T_{x,y}$
       $$
z\mapsto T_{x,y}(z)=T(x,y,z)
       $$
can be encoded by $2^r$ bits, so we let $a(x,y)=T_{x,y}$ and
$b(z)=z$ and get decomposition complexity at most $2^r+r$. The
bound $2^p+p$ is obtained in a symmetric way.

(Lower bound)~We can use a standard counting argument (in the same way
as in \cite{hansen-lachish-miltersen}; they consider functions, 
not predicates, but
this does not matter much.) 
Let us count how many possibilities we have for a predicate with
decomposition complexity $m$ or less. Choosing such a predicate,
we first have to choose numbers $u$ and $v$ such that $u+v\le m$.
Without loss of generality we may assume that $u+v=m$ (adding
dummy bits). First, let us count (for fixed $u$ and $v$)
all the decompositions where
$a$ has $u$-bit values and $b$ has $v$-bit values. 
We have $(2^u)^{2^{p+q}}$ possible $a$'s, $(2^v)^{2^{q+r}}$
possible $b$'s and $2^{2^{u+v}}$ possible $t$'s, i.e.,
       $$
2^{u2^{p+q}}\cdot
2^{v2^{q+r}}\cdot
2^{2^{u+v}} = 2^{u2^{p+q}+v2^{q+r}+2^{u+v}}\le 
2^{(u+v)2^{p+q}+(u+v)2^{q+r}+2^{u+v}}
       $$
possibilities (for fixed $u,v$). In total we get at most
       $$
m 2^{m2^{p+q}+
m2^{q+r}+
2^m}
       $$
predicates of decomposition complexity $m$ or less (the factor
$m$ appears since there are at most $m$ decompositions of $m$
into a sum of positive integers $u$ and $v$). Therefore, if all $2^{2^n}$ predicates
$\mathbb{B}^p\times\mathbb{B}^q\times\mathbb{B}^r\to\mathbb{B}$
have decomposition complexity at most $m$, then 
       $$
m 2^{m2^{p+q}+m2^{q+r}+2^m} \ge 2^{2^n}
       $$
or
       $$
\log m + m2^{p+q}+m2^{q+r}+2^m \ge 2^n
       $$
At least one of the terms in the left-hand side should be
$\Omega(2^n)$, therefore either $m\ge n-O(1)$ [if $2^m=\Omega(2^n)$], 
or $\log m \ge r-O(1)$ [if $m2^{p+q}\ge\Omega(2^n)=\Omega(2^{p+q+r})$], 
or $\log m \ge p-O(1)$
[if $m2^{q+r}\ge\Omega(2^n)=\Omega(2^{p+q+r})$].\qed

\subsection{Bounds for explicit predicates}

As with circuit complexity, an interesting question is to
provide a lower bound for an explicit function; it is usually much
harder than proving the existence results. The following
statement provides a lower bound for a simple function.

Consider the predicate $T\colon
\mathbb{B}^k\times\mathbb{B}^{2^{2k}}\times\mathbb{B}^k\to
\mathbb{B}$ defined
as follows:
       $$
T(x,y,z)=y(x,z)
       $$
where $y\in \mathbb{B}^{2^{2k}}$ is treated as a function
$\mathbb{B}^k\times\mathbb{B}^k\to \mathbb{B}$.

\begin{theorem}
\label{indexing}

The decomposition complexity of $T$
is at least $2^k$.
\end{theorem}

(Note that this lower bound almost matches the second upper bound
of Theorem~\ref{th:general}, which is $k+2^k$.)

\proof Assume that some decomposition of $T$ is given:
       $$
T(x,y,z)=t(a(x,y),b(y,z)),
       $$
where $a(x,y)$ and $b(y,z)$ consist of $u$ and $v$ bits respectively.
Then every $y:\mathbb{B}^k\times\mathbb{B}^k\to\mathbb{B}$
determines two functions $a_y\colon\mathbb{B}^k\to\mathbb{B}^u$
and $b_y\colon\mathbb{B}^k\to\mathbb{B}^v$ obtained from $a$ and
$b$ by fixing $y$. Knowing these two functions (and $t$) one should be
able to reconstruct $T(x,y,z)$ for all $x$ and $z$, since
       $$
T(x,y,z)= t(a_y(x),b_y(z)),
       $$
i.e., to reconstruct $y$. Therefore, the number of possible
pairs $\langle a_y,b_y\rangle$, which is at most
       $$
2^{u2^k} \cdot 2^{v2^k},
       $$
is at least the number of all $y$'s, i.e. $2^{2^{2k}}$. So we get
       $$
 (u+v)2^k \ge 2^{2k},
       $$
or $u+v\ge 2^k$, therefore the decomposition complexity of $T$
is at least $2^k$.\qed

\textbf{Remarks}.

\textbf{1}.
In this way we get a lower bound
$\Omega(\sqrt{n})$ (where $n$ is the total input size) for the case
when $x$ and $z$ are of size about $\frac{1}{2}\log n$.  In this
case this lower bound matches the upper bound of Theorem~\ref{th:general},
as we have noted.

\textbf{2}.
Here is another example where upper and lower bounds match.
If the predicate $t(x,y,z)$ is defined as $x=z$, we
need to transmit $x$ and $z$ completely (see \cite{nisan} or use the 
pigeon-hole principle).
So there is a trivial (and tight) linear lower bound if we let $x$
and $z$ be long (of $\Theta(n)$) size. 

\textbf{3}. It would be interesting
to get a linear bound for an explicit function in an 
intermediate case when $x$ and $z$ are short compared to $y$
(preferable even of logarithmic size) but not as short as in
Theorem~\ref{indexing} (so a non-constructive
lower bound applies).
Such a lower bound would mean that $a(x,y)$ or $b(y, z)$ has to
retain a significant part of information in $y$. Intuitive
explanation for this necessity could be: ``since we do not know $z$
when computing $a(x,y)$, we do not know which part of $y$-information
is relevant and need to retain a significant fraction of $y$''.
Note that for the function $T$ defined above this is not the case:
not knowing $z$, we still know $x$ so only one row ($x$th row)
in the matrix
$y$ is relevant.

The natural candidate is the function
       $
T'\colon \mathbb{B}^k\times \mathbb{B}^{2^k}\times\mathbb{B}^k\to\mathbb{B}
       $
defined by $T'(x,y,z)=y(x\oplus z)$.  Here $y$ is considered as a vector
$\mathbb{B}^k\to\mathbb{B}$,
not matrix, and $x\oplus z$ denotes bitwise XOR of two $k$-bit
strings $x$ and $z$. The size of $x$ and $z$ is about $\log n$ (where $n$ is the
total input size), and for these input sizes the worst-case lower bound is
indeed linear. One could think that this lower bound could be obtained for
$T'$: ``when computing $a(x,y)$ we do not know $z$, and $x\oplus z$
could be any bit string of length $k$, so all the information in $y$ is
relevant''. However, this intuition is false, and there exists a sublinear 
upper bound $O(n^{0.92})$, see~\cite{babai} or~\cite{nisan}, p.~95.\footnote{This upper
bound is obtained as follows.  Let us consider $y$ as a Boolean function of
$k$ Boolean variables; $y\colon (u_1,\ldots,u_k)\mapsto y(u_1,\ldots,u_k)$. 
Such a Boolean function can be represented as a multi-linear
polynomial of degree $k$ over the $2$-element field $\mathbb{F}_2$. This polynomial 
$y(u_1,\ldots,u_k)$ 
has $2^k$ bit coefficients
and is known when $a(x,y)$ or $b(y,z)$ are computed. Let us separate terms
of ``high'' and ``low'' degree in this polynomial: 
	$$
y(u_1,\ldots)=y_\textrm{low}(u_1,\ldots)+y_\textrm{high}(u_1,\ldots),
	$$
taking $\frac{2}{3}k$ as the threshold between ``low'' and ``high''. 
The polynomial $y_\textrm{high}$ is
included in $a$ (or $b$) as is, just by listing all its coefficients. (We have about
$2^{H(\frac{2}{3})k}\approx n^{0.92}$ of them, where $H$ is Shannon entropy
function.) For $y_\textrm{low}$ we use the following trick. Consider
$y(X_1\oplus Z_1,\ldots,X_k\oplus Z_k)$ as a polynomial $\tilde y$ of $2k$ variables $X_1,\ldots,X_k,Z_1,\ldots,Z_k\in\mathbb{F}_2$. Its degree is at most $\frac{2}{3}k$, and each
monomial includes at most $\frac{2}{3}k$ variables. So we can split $\tilde y$ again:
	$$
\tilde y(X_1,\ldots,Z_1,\ldots)=\tilde y_{x\textrm{-low}}(X_1,\ldots,Z_1,\ldots) +\tilde y_{z\textrm{-low}}(X_1,\ldots,Z_1,\ldots);
	$$
here the first term has small $X$-degree ($Z$-variables are treated as constants),
and the second term has small $Z$-degree. Here ``small'' means ``at most $\frac{1}{3}k$''.
All this could be done 
in both nodes (while computing $a$ and $b$), since $y$ is known there;
$X_i$ and $Z_i$ are just variables. Now we include in $a(x,y)$ the
coefficients of the polynomial $(Z_1,\ldots,Z_k)\mapsto\tilde y_{z\textrm{-low}}
(x_1,\ldots,x_k,Z_1,\ldots,Z_k)$, and do the symmetric thing for $b(y,z)$.
Both polynomial have degree at most $\frac{1}{3}k$, so we again
need only $O(n^{0.92})$ bits to specify them.}
(This upper bound should be compared to
the $\Omega(\sqrt{n})$ lower bound obtained
by reduction to $T$: in the special case when the left half of $x$ and the
right half of $z$ contain only zeros, we get $T$ out of $T'$.)

\textbf{Question}: what happens if we replace $x\oplus z$ by $x+z\bmod 2^k$
in the definition of $T'$? It seems that the upper bound argument does not
work any more.

\section{Probabilistic decomposition}

As in communication complexity theory, we may consider also
probabilistic and distributional versions of decomposition
complexity. In the probabilistic version we consider random variables
instead of binary functions $a, b, t$ (with shared random bits or
independent random bits). In the distributional version we
look for a decomposition that is
Hamming-close to a given function. 

It turns out that the lower bounds mentioned above are robust in that
sense and remain valid for distributional (and therefore probabilistic)
decomposition complexity almost unchanged.

Let $\varepsilon$ be a positive number less than $1/2$. We are
interested in a minimum decomposition complexity of a function
that $\varepsilon$-approximates a given one (coincides with it
with probability at least $1-\varepsilon$ with respect to uniform distribution
on inputs). For $\varepsilon\ge \frac{1}{2}$ this question is trivial 
(either $0$ or $1$ constant provide the required approximation).
So we assume that some $\varepsilon<\frac{1}{2}$ is fixed (the $O()$-constants
in the statements will depend on it). 

A standard argument shows that lower bounds established for 
distributional decomposition complexity  remain
true for probabilistic complexity (where $a,b,t$ use random bits and 
for every input $x,y,z$ the random variable $t(a(x,y),b(y,z))$ should
coincide with a given function with probability at least $1-\varepsilon$).
So we may consider only the distributional complexity.

\begin{theorem}\label{th:prob}
\textbf{\textup{(1)}}~Let $n=p+q+r$ and $p,r\ge \log n+O(1)$. Then there 
exists a predicate
$T\colon \mathbb{B}^p\times\mathbb{B}^q\times\mathbb{B}^r\to \mathbb{B}$
such that decomposition complexity of any its $\varepsilon$-approximation
is at least $n-O(1)$.

\textbf{\textup{(2)}}~For the predicate $T$ used in Theorem~\ref{indexing}
we get the lower bound $\Omega(2^k)$ \textup(in the same setting\textup).
\end{theorem}

\proof \textbf{1}. Assume this is not the case. We repeat the same
counting argument as in Theorem~\ref{th:general}. Now we have
to count not only the predicates that have decomposition
complexity at most $m$, but also their $\varepsilon$-approximations.
The volume of an $\varepsilon$-ball in $\mathbb{B}^{2^{n}}$ is
about $2^{H(\varepsilon)2^{n}}$, so the number of the centers
of the balls that cover the entire space is at least 
$2^{(1-H(\varepsilon))2^{n}}$. So  after taking the logarithms
we get a constant factor $(1-H(\varepsilon))$, and the lower
bound for $m$ remains $n-O(1)$.

\textbf{2}. If the computation is correct for $1-\varepsilon$ fraction 
of all triples $(x,y,z)$, then there exist $\varepsilon'<\frac{1}{2}$ and
$\varepsilon''>0$ such that for at least $\varepsilon''$-fraction of 
all $y$ the computation is correct with probability at least $1-\varepsilon'$
(with respect to uniform distribution on $x$ and $z$).
This means that $\varepsilon'$-balls around functions 
$(x,z)\mapsto t(a_y(x),b_y(z))$ cover at least $\varepsilon''$-fraction
of all functions $y$. (See the proof of Theorem~\ref{indexing}.)
Again this gives us a constant factor before $2^{2k}$, but here we
do not take the logarithm second time, so we get $u+v\ge\Omega(2^k)$,
not $2^k-O(1)$.
\qed

\section{Applications to cellular automata}

An (one-dimensional) cellular automata is a linear array of
cells. Each of the cells can be in some state from a finite set
$S$ of states (the same for all cells). At each step all the
cells update their state; new state of a cell is some fixed function of
its old state and the states of its two neighbors. All the updates
are made synchronously.

Using a cellular automaton to compute a predicate, we assume
that there are two special states $0$ and $1$ and a neutral
state that is stable (if a cell and both its neighbors
are in the neutral state, then the cell remains neutral). To compute $P(x)$ for a
$n$-bit string $x$, we assemble $n$ cells and put them into
states that correspond to $x$; the rest of the (biinfinite) cell
array is in a neutral state.

Then we start the computation; the answer should appear in some
predefined cell (see below about the choice of this cell).

There is a natural non-uniform version of cellular automata: we
assume that in each vertex of the time-space diagram an
arbitrary ternary transition function (different for different
vertices) is used. Then the only restriction is caused
by the limited capacity of links: we require that
inputs/outputs of all functions (in all vertices) belong to some
fixed set $S$.

In this non-uniform setting a predicate $P$ on binary strings is
considered as a family of Boolean functions $P_n$ (where
$P_n$ is a restriction of $P$ onto $n$-bit strings) and for each
$P_n$ we measure the minimal size of a set $S$ needed to compute
$P_n$ in a non-uniform way described above. If this size is an
unbounded function of $n$, we conclude that predicate $P$ is not
computable by a cellular automaton. (In classical complexity theory we use
the same approach when we try to prove that some predicate is
not in P since it needs superpolynomial circuits in a non-uniform
setting.)

As usual, getting lower bounds for nonuniform models is
difficult, but it turns out that decomposition complexity
can be used if the cellular automaton is required to produce the
answer as soon as possible.

Since each cell gets information only from 
itself and its two neighbors, the first occasion to use all
$n$ input bits happens around time $n/2$ in the middle of the
string:
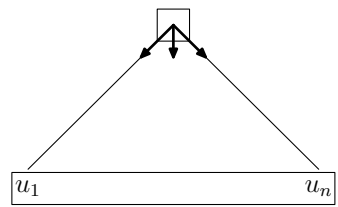
\begin{center}
       \includegraphics[scale=0.85]{decomp-7.mps}
\end{center}

Now we assume that the output of a cellular automaton is
produced at this place (both in uniform and non-uniform model).
(This is a very strong version of real-time computation by
cellular automata; we could call it ``as soon as
possible''-computation.)

The next theorem observes that non-uniformly computable family
of predicates is transformed into a function with small decomposition
complexity if we split the input string in three parts. 

\begin{theorem}
Let $T_k \colon \mathbb{B}^{k+f(k)+k}=
\mathbb{B}^k\times \mathbb{B}^{f(k)} \times
\mathbb{B}^k\to \mathbb{B}$ be a family of predicates that is
non-uniformly computable in this sense. Then the decomposition
complexity of $T_k$ is $O(k)$, and the constant in $O$-notation
is the logarithm of the number of states.
\end{theorem}

\begin{figure}[h]       
\begin{center}
       \includegraphics[scale=0.85]{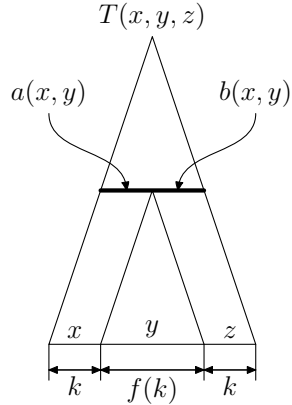}
\end{center}
\caption{Automaton run and its decomposition.}\label{decomp-automaton}
\end{figure}

\proof Consider Figure~\ref{decomp-automaton} where
the (nonuniform) computation is presented
(we use bigger units for time direction to make
the picture more clear).

Let us look at the contents of the line of length $2k$ located $k$
steps before the end of the computation. The left half is
$a(x,y)$, the right half is $b(y,z)$ and the function $t$ is
computed by the upper part of the circuit. It is easy to see
that $a(x,y)$ indeed depends only on $x$ and $y$ since
information about $z$ has not arrived yet; for the same reason
$b(y,z)$ depends only on $y$ and $z$. The bit size of $a(x,y)$
and $b(y,z)$ is $k\log\#S$.\qed


\begin{corollary}The predicate $T$ from
Theorem~\ref{indexing} cannot be computed in this model.
\end{corollary}

This predicate splits a string of length $k+2^{2k}+k$
into three pieces $x,y,z$ of length $k$, $2^{2k}$ and $k$ respectively,
and then computes $y(x,z)$. Note that this can be done 
by a cellular automaton
in linear time. Indeed, we combine the string $x$ and $z$ into a
$2k$-binary string; then we move this string across the middle
part of input subtracting one at each step and waiting until our
counter decreases to zero; then we know where the output bit
should be read. So we get the following result:

\begin{theorem}
       \label{separation}
There exists a linear-time computable predicate that is not
computable ``as soon as possible'' even in a non-uniform model.
\end{theorem}

\textbf{Remark}. This result and the intuition behind the proof
are not new (see the paper of V.~Terrier~\cite{terrier}; see
also~\cite{culik}). However, the explicit use of decomposition
complexity helps to formalize the intuition behind the proof. It
also allows us to show (in a similar way) that this predicate
cannot be computed not only ``as soon as possible'', but even
after $o(\sqrt{n})$ steps after this moment (which seems to be
an improvement).

Another improvement that we get for free is that we cannot even
$\varepsilon$-approximate this predicate in the ``as soon as
possible'' model.

\textbf{Question}: There could be other ways to get lower
bounds for non-uniform automata (=triangle circuits). Of course,
there is a counting lower bound, but this does not give any
explicit function. Are there some other tools?

\section{Algorithmic Information Theory}\label{sec:multisource}

Now we can consider the Kolmogorov complexity version of the
same decomposition problem. Let us start
with some informal comments. Assume that we have four binary
strings $x,y,z,t$ such that $\KS(t|x,y,z)$ is small
(we write $\KS(t|x,y,z)\approx 0$, not specifying exactly how 
small should it be).  Here
$\KS(\alpha|\beta)$ stands for conditional complexity of
$\alpha$ when $\beta$ is known, i.e., for the minimal length of
a program that transforms $\beta$ to $\alpha$. (Hence our
requirement says that there is a short program that produces $t$
given $x,y,z$.)

We are looking for strings $a$ and $b$ such that
$\KS(a|x,y)\approx 0$, $\KS(b|y,z)\approx 0$, and
$\KS(t|a,b)\approx 0$. Such $a$ and $b$ always exist, since we
may let $a=\langle x,y\rangle$ and $b=\langle y,z\rangle$
(again, $y$ can also be split between $a$ and $b$). However, the
situation changes if we restrict the complexities of $a$ and $b$ (or their
lengths, this does not matter, since each string can be replaced
by its shortest description).
As we shall see, sometimes we need $a$ and $b$ of total
complexity close to $\KS(x)+\KS(y)+\KS(z)$ even if $t$ has much
smaller complexity. (Note that now we cannot restrict ourselves to one-bit
strings $t$ for evident reasons.)

To be specific, let us agree that all the strings $x,y,z,t$ have the
same length $n$; we look for strings $a$ and $b$ of length
$m$, and ``small'' conditional complexity means that complexity
is less than some $c$.

\begin{theorem}
     \label{th:complexity-1}
If $3c<n-O(1)$ and $2m+c<3n-O(1)$, there exist 
strings $x,y,z,t$ of length $n$ such that $K(t|x,y,z)=O(\log n)$,
but there are no strings $a,b$ of length $m$ such that
	$$
K(a|x,y)<c,\qquad K(b|y,z)<c, \qquad K(t|a,b)<c.	
	$$
\end{theorem}

For example, this is true if $c=O(\log n)$ and $m$ is $1.5n-O(\log n)$ 
(note that for $m=1.5n$ we can split $y$ into two halves and combine
the first half with $x$, and the second half with $y$).

\proof Consider the following algorithm. Given $n$, we generate
(in parallel for all $x,y\in\mathbb{B}^n$) the lists of those $m$-bit strings who
have conditional complexity (with respect to $x$ and $y$) less than $c$ (one list
for each pair $x,y$).
Also we generate (in parallel for all strings $a$ and $b$ of length $m$)
the lists of those strings $t$ who have complexity less than $c$ given $a$ and $b$ 
(one list for each pair $a,b$).
At every step of enumeration we imagine that these lists are final and construct
a quadruple
$x,y,z,t$ that satisfies the statement of the theorem.  It is done as
follows: we take a ``fresh'' triple $x,y,z$ (that was not used on the
previous steps of the construction), take all strings $a$ that are in
the list for $x,y$, take all strings $b$ that are in the list for $y,z$, and take
all strings $t$ that are in the lists for those $a$s and $b$s. Then we choose
some $t$ that does not appear in all these lists.

Such a $t$ exists since we have at most $2^c$ strings $a$ (for given $x$ and $y$),
and at most $2^c$ strings $b$ (for given $y$ and $z$). For every of $2^{2c}$ pairs 
$(a,b)$ there are at most $2^c$ strings $t$, so in total at most $2^{3c}$ values 
of $t$ are unsuitable, and we can choose a suitable one.

We also need to ensure that there are enough ``fresh'' pairs for
all the steps of the construction. The new elements
in the first series of lists may appear at 
most $2^{n}\times 2^{n}\times 2^c$ times (we have
at most $2^n\times 2^n$ pairs $(x,y)$ and at most $2^c$ values of $a$ for
each pair). Then we have $2^{m}\times 2^{m}\times 2^{c}$ events for the
second series of lists. On the other hand, we have $2^{3n}$ triples $(x,y,z)$,
so we need the inequality
	$$
2^{2n+c} + 2^{2m+c}< 2^{3n},
	$$
which is guaranteed by our assumptions.

To run this process, it is enough to know $n$, so for every $x,y,z,t$ generated
by this algorithm we have $K(t|x,y,z)=O(\log n)$. (For given $x,y,z$ only one $t$
may appear since we take a fresh triple each time.)
\qed

This result can be improved: 

\begin{theorem}
Assume that $3c<n-O(1)$ and $m\le 1.5n-O(\log n)$.  
We can effectively construct for every $n$ a
total function $T:\mathbb{B}^n\times \mathbb{B}^n\times\mathbb{B}^n\to
\mathbb{B}^n$ such that for random 
\textup(=\,incompressible\textup) triple $x,y,z$ and
$t=T(x,y,z)$ the strings $a$ and $b$ of length $m$ 
that provide a decomposition
\textup(as defined above\textup) do
not exist.
\end{theorem}

The improvement is two-fold: first, we have a total function $T$ (instead of
a partial one provided by the previous construction); second, we claim that
all random triples have the required property (instead of mere existence of
such a triple). 

\proof Let us first deal with the first improvement.
Consider multi-valued functions $A,B\colon \mathbb{B}^n\times
\mathbb{B}^n\to\mathcal{P}(\mathbb{B}^m)$ that map every pair of $n$-bit
strings into a $2^c$-element set of $m$-bit strings. Consider also
multi-valued function $F\colon \mathbb{B}^m\times\mathbb{B}^m\to\
\mathcal{P}(\mathbb{B}^n)$ whose values are $2^c$-element sets of $n$-bit
strings. We say that $A,B,F$ \emph{cover} a total function 
$T:\mathbb{B}^n\times \mathbb{B}^n\times\mathbb{B}^n\to
\mathbb{B}^n$ if for every $x,y,z\in\mathbb{B}^n$ there exist strings
$a,b\in\mathbb{B}^m$ such that $a\in A(x,y)$, $b\in B(y,z)$, and 
$T(x,y,z)\in F(a,b)$. 

Let us prove first the following combinatorial statement: \emph{there exists
a function $T$ that is not covered by any triple 
of functions $A,B,F$}. This can be shown
by a counting argument similar to the proof of Theorem~\ref{th:general}.
Indeed, let us compute the probability of the event ``random function $T$
is covered by some fixed $A,B,F$''. This event is the intersection of
independent events (for each triple $x,y,z$). For given $x,y,z$ there are
$2^c$ possible $a$s, $2^c$ possible $b$s, and $2^c$ possible elements
in $F(a,b)$ for each $a$ and $b$, i.e., $2^{3c}$ possibilities altogether.
Since $3c<n-O(1)$, each of the independent events has probability less
than $\frac{1}{2}$, and their intersection has probability less than 
$2^{-2^{3n}}$.

This probability then should be multiplied by the number of triples $A,B,F$.
For $A$ and $B$  we have at most $(2^m)^{2^n\times 2^n\times 2^c}$ possibilities,
for $F$ we have at most $(2^n)^{2^m\times 2^m\times 2^c}$ possibilities.
So the existence of a function $T$ not covered by any triple is guaranteed if
	$$
2^{m2^{2n+c}}\times 2^{m2^{2n+c}}\times 2^{n2^{2m+c}}\times 2^{-2^{3n}}<1,
	$$
i.e.,
	$$
m2^{2n+c}+m2^{2n+c}+n2^{2m+c} < 2^{3n},
	$$
and this inequality follows from the assumptions.

The property ``$T$ can be covered by some triple $A,B,F$'' can be computably
tested by an exhaustive search over all triples $A,B,F$. So we can 
(for every $n$) computably find the first (in some order) function $T$ that
does not have this property. For these $T$ there are some $x,y,z$ that do not
allow decomposition. Indeed, we can choose $A$ so that
$A(x,y)$ contains all strings $a$ of length $m$ such that $K(a|x,y)<c$, etc.

However, we promised more: we need to show not only the existence of $x,y,z$
but that all incompressible triples (this means that $K(x,y,z)\ge 3n-O(1)$) have
the required property. This is done in two steps. First, we show than (for
some $F$ that computably depends on $n$)
most triples do not allow decomposition. Then we note that one can enumerate
triples that allow decomposition, so they can be encoded by their
ordinal number in the enumeration and therefore are compressible.

To make this plan work, we need to consider other property of function $T$.
Now we say that $T$ is covered by $A,B,F$ if at least $2^{-O(1)}$-fraction of
all triples $(x,y,z)$ admit $a$ and $b$. The probability of this event should
now be estimated by Chernoff inequality (we guarantee first that the probability of
each individual event  is, say, twice smaller than the threshold), and we get
a bound of the same type, with $\Omega(2^{3n})$ instead of $2^{3n}$, which
is enough.\qed

In fact, this argument provides a decomposition complexity bound similar
to Theorem~\ref{th:general}, but now the functions $a$, $b$ and $t$ are 
multi-valued and we can choose any of their values to obtain $t(x,y,z)$.

\subsection*{Remarks and questions}

\textbf{1}. Similar results can be obtained for more binary operations in the
decomposition. Imagine that we have some strings $x,y,z,t$ of length $n$ such
that $K(t|x,y,z)$ is small and want to construct some ``intermediate''
strings $u_1,\ldots,u_s$ such that in the sequence
	$$
x,y,z,u_1,u_2,\ldots,u_s,t
	$$
every string, starting from $u_1$, is conditionally simple with 
respect to some \emph{pair} of  its
predecessors. We can use our technique to show that this is not possible
if all $u_i$ have length close to $n$ and the number $s$ is not large. 

\textbf{2}. As before, it would be nice to get lower bounds for some explicit
function $T(x,y,z)$ (even a non-optimal lower bound, 
like in Theorem~\ref{indexing}) for
the algorithmic information theory version of decomposition problem.

\textbf{3}. Many results of multi-source algorithmic information theory have some 
counterparts in classical information theory. Can we find some statement
that corresponds to the lower bound for decomposition complexity?

\textbf{4}. Is it possible to use the techniques of \cite{hansen-lachish-miltersen}
to get some bounds for explicit functions in algorithmic information
theory setting?

\end{document}